\documentclass[sigconf]{acmart}

\newtheorem{myDef}{Definition}
\usepackage{booktabs} %
\usepackage{bm}
\usepackage{multirow}
\usepackage{xspace}
\usepackage[tight,footnotesize]{subfigure}
\usepackage{algorithm}  
\usepackage{algpseudocode}  
\usepackage{amsmath}
\usepackage{amsfonts}
\usepackage{graphicx}
\usepackage{enumitem}
\usepackage{verbatim}
\usepackage{balance}

\newcommand{\methodshort}{\textbf{MotifGNN}}
\newcommand{\eg}{\emph{e.g.,}\xspace}

\AtBeginDocument{%
  \providecommand\BibTeX{{%
    \normalfont B\kern-0.5em{\scshape i\kern-0.25em b}\kern-0.8em\TeX}}}

\copyrightyear{2023}
\acmYear{2023}
\setcopyright{acmlicensed}\acmConference[KDD '23]{Proceedings of the 29th ACM SIGKDD Conference on Knowledge Discovery and Data Mining}{August 6--10, 2023}{Long Beach, CA, USA} \acmBooktitle{Proceedings of the 29th ACM SIGKDD Conference on Knowledge Discovery and Data Mining (KDD '23), August 6--10, 2023, Long Beach, CA, USA} \acmPrice{15.00} \acmDOI{10.1145/3580305.3599351} \acmISBN{979-8-4007-0103-0/23/08}

\begin{document}

\title{Financial Default Prediction via  Motif-preserving Graph Neural Network with Curriculum Learning}

\author{Daixin Wang}
\email{daixin.wdx@antgroup.com}
\affiliation{
  \institution{Ant Group}
  \country{China}
}
\author{Zhiqiang Zhang}
\email{lingyao.zzq@antgroup.com}
\affiliation{
  \institution{Ant Group}
  \country{China}
}
\author{Yeyu Zhao}
\email{yeyu.zyy@antgroup.com}
\affiliation{
  \institution{Ant Group}
  \country{China}
}
\author{Kai Huang}
\email{kevin.hk@antgroup.com}
\affiliation{
  \institution{Ant Group}
  \country{China}
}
\author{Yulin Kang}
\email{yulin.kyl@antgroup.com}
\affiliation{
  \institution{Ant Group}
  \country{China}
}
\author{Jun Zhou}
\email{jun.zhoujun@antgroup.com}
\authornote{Corresponding Author.}
\affiliation{
  \institution{Ant Group}
  \country{China}
}
\renewcommand{\shortauthors}{Daixin Wang et al.} %

\begin{abstract}
User financial default prediction plays a critical role in credit risk forecasting and management. It aims at predicting the probability that the user will fail to make the repayments in the future. Previous methods mainly extract a set of user individual features regarding his own profiles and behaviors and build a binary-classification model to make default predictions. However, these methods cannot get satisfied results, especially for users with limited information. Although recent efforts suggest that default prediction can be improved by social relations, they fail to capture the higher-order topology structure at the level of small subgraph patterns. In this paper, we fill in this gap by proposing a motif-preserving Graph Neural Network with curriculum learning (\methodshort) to jointly learn the lower-order structures from the original graph and higher-order structures from multi-view motif-based graphs for financial default prediction. Specifically, to solve the problem of weak connectivity in motif-based graphs, we design the motif-based gating mechanism. It utilizes the information learned from the original graph with good connectivity to strengthen the learning of the higher-order structure. And considering that the motif patterns of different samples are highly unbalanced, we propose a curriculum learning mechanism on the whole learning process to more focus on the samples with uncommon motif distributions. Extensive experiments on one public dataset and two industrial datasets all demonstrate the effectiveness of our proposed method.
\end{abstract}




\maketitle
\section{Introduction}
Recent years have witnessed a growing trend in internet financial services. For example, in China, Alipay and WechatPay all announced that they had more than 1 billion daily active users\footnote{https://www.merchantsavvy.co.uk/mobile-payment-stats-trends/}. Such a huge number of users facilitate many online credit services like credit pay, personal cash loan and enterprise cash loan. The basic idea of these services is that users can use their own credit as a pledge to obtain the limits to do payments or get cash. And after the user repays the money within the time stipulated in the contract, their credit limits will be recovered. 

User default prediction is at the heart of credit services. The purpose of the default prediction is to predict whether the user will fail to repay the money in the future. If the user can repay the money on time, it is labeled as a normal user, otherwise a default user. Therefore, it is commonly regarded as a binary classification problem. And early works mainly adopt some tree-based \cite{randhawa2018credit} or neural network-based models \cite{fiore2019using} on users' individual features from various aspects, like user profiles \cite{randhawa2018credit} and previous loan history \cite{yue2007review}. However, unlike traditional loan services, in online credit services, users are not required to provide pledge or previous loan information. And most users are new and have no previous records in the online platform, \eg $60$\% of the users have no previous loan records in PPDai, an online lending platform in China \cite{yang2019understanding}. In this way, users', especially new users' own features are very sparse, which poses a great challenge for previous methods to work well. 

\begin{figure}[htb]
\centering
\includegraphics[width=0.45\textwidth]{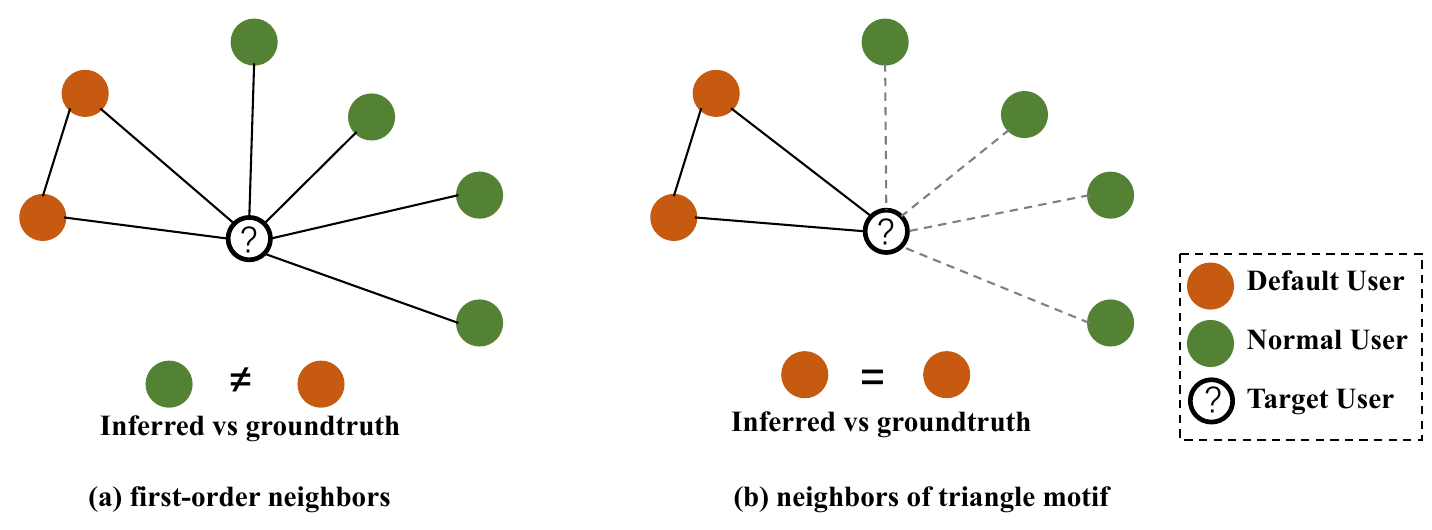}
\caption{ A real case in an industrial dataset. (a) Only using the target user's first-order neighbors may result in a wrong inference for the target user. (b) When we limit his neighbors using the triangle motif, we can correctly infer the target user as a default user. }
\label{pic:case_motif}
\end{figure}

In online financial platforms, besides applying for loans, users usually can do transfers, do transactions or add friends with each other, which also poses another possibility of improvement for financial default prediction. Then following works start to make efforts on exploiting the user's interactive relations as an information supplement for detecting the financial anomalies, e.g., default, fraud and cash-out. These works mainly use the social graphs \cite{wang2019semi}, money transfer graphs, transaction graphs \cite{hu2019cash} or their combinations as heterogeneous graphs to correlate users \cite{zhong2020financial}. Then these works propose homogeneous graph neural network (GNN) models \cite{wang2019semi}, heterogeneous GNN models \cite{hu2019cash,zhong2020financial} or spatial-temporal based GNN models \cite{yang2021financial,wang2021temporal} for financial anomaly prediction. These methods all hinge on a layer-wise propagation mechanism based on the lower-order graph structure at the level of individual nodes and edges, e.g. first-order and second-order neighbors. Despite of their effectiveness, previous models overlook the higher-order structure patterns at the level of small sub-graph patterns and may get sub-optimal results. Figure \ref{pic:case_motif} gives a real case in a real-world dataset: The first-order neighbors of the target user are mostly the normal users. Previous methods are likely to mistakenly infer the target user as the normal user, since they only model the direct edges. But if we consider the higher-order relations, we can find that the target user and the two default users form a triangle relation, which indicates a stronger bond and may help us infer the target user as a default user. Therefore, it is essential to go beyond the lower-order structure and incorporate the higher-order structure \cite{rossi2018estimation,rossi2018higher,carranza2018higher}. 

Network motifs have been demonstrated to be able to characterize the higher-order structure of the graph. They are small graphs involving the coupling effect of a small group of nodes and edges, like the triangle subgraph in Figure \ref{pic:case_motif}(b). By preserving the motifs of each node, the model is likely to capture the higher-order connectivity patterns for each node and thus give a more accurate depiction of the entire graph structure \cite{benson2016higher,arenas2008motif,lyu2017enhancing}. Although some research works have tried to incorporate motifs into GNN models \cite{lee2019graph,wang2022graph,piao2021predicting}, they still suffer from the following two challenges: 

(1) \textbf{Weak Connectivity}: These motif-based GNN models assume that the motif-based graph has as good connectivity as many graphs. Thus they typically focus only on the higher-order structures that are encoded in the motif-based graph, but underestimate or even violate the original lower-order connections. However, it is worth noting that two nodes can be separated in higher-order connections but are connected in the original graph, thereby the graph built upon the motifs may be fragmented into several connected components or even isolated nodes. Such weak connectivity poses a great challenge for existing models to work well. 

(2) \textbf{Motif Patterns Unbalance}: The higher-order structures carried by different motifs are highly unbalanced in real-world graphs. For example, the number of edges in some motif-based graphs may be thousands times as those of other types of motifs. Therefore, the structures of different users have diverse and unbalanced distributions on motif patterns. In this way, some nodes will have uncommon motif-based patterns, which may be underestimated and are difficult to be learned effectively by previous methods.

To address these challenges, we propose a motif-preserving Graph Neural Network with Curriculum Learning (\methodshort), where we jointly model the coupling effect of lower-order and higher-order graph structure in social graphs for default prediction. Specifically, (1) we design a multi-view graph encoder with motif-based gates to learn from the original graph and multi-view motif-based graphs. In particular, considering the weak connectivity of motif-based graphs, the designed gating mechanism can assist the learning of the higher-order structural information by the representation learned from the original graph with good connectivity. (2) Secondly, we define a term to measure the deviation of the user's motif pattern distributions from the averaged distributions, and accordingly design a curriculum learning mechanism to make the learning process progressively focus on the users with uncommon patterns. We evaluate our method on a public dataset and two real-world industrial datasets. The results on all the datasets demonstrate the superiority of our method. 

The contributions of this work can be summarized as follows:
\begin{itemize}
    \item \textbf{Problem}: To the best of our knowledge, we are the first to leverage both the lower-order and higher-order social graph structures into the problem of financial default prediction. 
    \item \textbf{Methodology}: We propose a novel motif-preserving graph neural network method with curriculum learning to address the weak connectivity and motif pattern unbalanced issues when incorporating the motifs into graph models.
    \item \textbf{Result}: Our experimental results on a public dataset and two industrial datasets all demonstrate that our method achieves the best performance among the state-of-the-art methods. 
\end{itemize}

\section{Related Work}
\subsection{Default Prediction}
User Default Prediction is a very central task of credit risk management for both traditional credit card institutions like banks and newly emerging Internet financial platforms. Previous works are mainly from the field of economics and focusing on designing some rules to predict credit risk \cite{whitrow2009transaction,bermudez2008bayesian,viaene2007strategies}. However, manually defining rules is easily to be attacked and difficult to deal with the evolving fraud patterns. Then some methods from the data mining field are emerging. They mainly regard each user as an independent entity and feed the user's profiles and historical behaviors into supervised learning approaches like support vector machines, tree-based methods or neural networks \cite{fiore2019using,randhawa2018credit}. For example, Bhattacharyya et al. \cite{bhattacharyya2011data} analyzed several user profiles and derived $16$ important attributes like average amount over $3$ months to evaluate various methods like random forest and support vector machines for credit card fraud detection. Zhang et al. \cite{zhang2019distributed} collected more than $5000$ original features from different aspects like the transaction features, the historical features and profiles and used distributed deep forest model to detect cash-out users. However, users in online financial platforms are not active enough compared with e-commerce platforms, which makes their own features sparse and not informative enough. Therefore, only relying on the user's own features are difficult to model his default probability accurately. Afterward, a few recent works start to utilize different types of graphs like the social graph, the transaction graph, the user-device graph and their combinations as heterogeneous graphs to link users and propose graph-based methods to extract the complex interactive patterns between users \cite{liu2019geniepath,wang2019semi,hu2019cash,zhong2020financial,liang2021credit}. For example, Liu et al. \cite{liu2018heterogeneous} utilized a user-device graph and proposed a graph neural network for malicious account detection. Hu et al. \cite{hu2019cash} and Zhong et al. \cite{zhong2020financial} proposed heterogeneous graph neural networks with meta-path and hierarchical attention  to mine the heterogeneous graph for default user prediction. Wang et al. \cite{wang2019semi} proposed a spatial-temporal model to detect both the structural and temporal signal for default user detection. However, all of these works only consider the direct edges between entities, but overlook the higher-order structure within the graph, which makes our work distinct from the existing works. 

\subsection{Graph Neural Network}
Recently, graph neural network (GNN) has achieved state-of-the-art performance for data generated from non-Euclidean domains and represented as graphs \cite{zhang2020deep,wu2020comprehensive}. Earlier works focus on GNN models on spectral domains \cite{cnn_graph,chen201fastgcn}, which rely on the specific eigenfunctions of the Laplacian Matrix. Then following works focus on GNN models on spatial domains like GraphSAGE \cite{hamilton2017inductive} and GAT \cite{velivckovic2017graph}, which define the aggregation functions directly based on the node neighbors. Afterward, the methods for more complex graphs are proposed, like MultiGCN for multi-view graphs \cite{Khan2019MultiGCNGC}, HAN for heterogeneous graphs \cite{han2019}, STGCN for spatial-temporal graphs \cite{yu2018spatio}. And most of these works are based on the lower-order graph structure at the level of individual nodes and edges for common machine learning scenarios like recommendation and social network analysis. We extend GNN models to the problem of financial default prediction, by capturing both the lower-order graph structure and the higher-order graph structure at the level of small subgraphs.
. 
\subsection{Network Motif}
Network motifs are regarded as basic building blocks of complex networks \cite{milo2002network} and the investigation of motifs has been demonstrated to play essential roles in many kinds of networks like biology networks \cite{prill2005dynamic} and social networks \cite{rotabi2017detecting}. Furthermore, different types of motifs possess different kinds of graph properties. For example, Milo et al. find that motifs of feed-forward loops act as information processing in neurons and gene regulation networks, and motifs of three chains can characterize the energy flows in food webs \cite{milo2002network}.

Inspired by the good properties of network motifs, many graph-based machine learning models including GNN-based models incorporate motifs to account for higher-order network structures \cite{ahmed2020role,huang2020motif,rossi2018higher,lee2019graph,wang2022graph,hu2022mbrep,liu2021motif,piao2021predicting,wang2017community}. For example,  Lee et al. defined the motif-based adjacency matrices and proposed a motif-based graph attention network to mine the high-order structure in these matrices \cite{lee2019graph}. Hu et al. defined some heterogeneous motifs and extended the network motifs into the heterogeneous graph neural network \cite{hu2022mbrep}. Liu et al. proposed a motif-preserving temporal shift based GNN model to simultaneously model the local high-order structures and temporal evolution for dynamic attributed networks \cite{liu2021motif}. All of these works directly model the higher-order topology structure via the motif-based graph, but underestimate or even violate the original lower-order connections. Actually, the motif-based graph has weak connectivity and different users have unbalanced motif patterns, which are still unsolved problems for all of these models.  

\section{Problem Formulation and Data Analysis}
In this section, we first introduce some notations and formally define our task of financial default prediction. Then we give some data analysis to articulate our motivation to incorporate both the lower-order and higher-order social relations into our problem. 

\subsection{Problem Formulation}
\label{sec:pf}
To properly formalize the task of incorporating the social graphs into financial default prediction, we first define our social graph $\mathcal{G}$: 

\begin{myDef}
(Social Graph) The Social Graph in my problem is represented as a directed graph $\mathcal{G}=(\mathcal{V},\mathcal{E},\mathbf{X})$, where $\mathcal{V}$ is the set of users with $|\mathcal{V}|=n$, and $\mathcal{E}$ is the set of user social relations with $|\mathcal{E}|=e$, $\mathbf{X}\in\mathbb{R}^{n\times d_{nd}}$ denotes the user features with the dimension of $d_{nd}$. And the adjacency matrix of the graph $\mathcal{G}$ is denoted as $\mathbf{A}\in \mathbb{R}^{n\times n}$.
\end{myDef}

Note that the user features are extracted from the user's profiles, the behavior history in our platform and the loan history. Specifically, the user profiles contain personal information, including gender, age, register time and et al. The behavior history represents some statistical features about the user's behaviors in our platform, like log-in frequency, online duration, click information and et al. The loan history contains the user's previous loan information. In addition, the social graph in our problem is constructed based on social relations between users, including friendship, trade and transfer relations.

Then our task of financial default prediction can be formulated as follows: 
\begin{myDef}
(Financial Default Prediction) Given the social graph $\mathcal{G}=(\mathcal{V},\mathcal{E},\mathbf{X})$, among which the node set contains the labeled users $\mathcal{V}_L$ and unlabeled users $\mathcal{V}_{uL}$, our problem aims to learn from the labeled data and its graph structure to predict the default probability on unlabeled users, defined as
\begin{equation}
    \hat{y}_u = \mathop{\arg\max}{P(y_u|\mathcal{G})}, u \in \mathcal{V} \ and \ y_u \in \{0,1\}\nonumber
\end{equation}
\end{myDef}

Generally speaking, the financial default prediction can be regarded as a binary node classification problem on a social graph. 

\subsection{Data Analysis}
Before directly modeling the social graph into the financial default prediction, we first give some data analysis to illustrate our hypothesis that both the lower-order and higher-order graph structures contribute to financial default prediction. 

To validate the hypothesis, we conduct empirical data analysis on a sampled industrial dataset, which consists of about $1$ million users with their social graph. And we have two default labels for these users: one label for the consumption loan product like credit card and the other label for the cash loan product.  
We first explore the ratio (lift) between the average bad rate of default users' neighbors and normal users' neighbors on the two labels. The result is shown in Table \ref{tab:data_analysis1}. We find that no matter for the direct neighbors or for the second-order neighbors, there exists a distinct lift between the default users' neighbors and normal users' neighbors: Default users' first-order and second-order neighbors have much larger probability to default compared with normal users' neighbors. It demonstrates that considering both the first-order and second-order neighbors of the users is effective for default user prediction. 

\begin{table}[htb]
\centering
\caption{ The ratio between the average bad rate of default users' neighbors and the bad rate of normal users' neighbors on two types of loans. $br(\mathcal{N}_{y}^{(l)})$ denotes the average bad rate of the $l$-order neighbors of users with label $y$. }
\begin{tabular}{|c|c|c||c|c|}
\hline
\multirow{2}{*}{Methods} &
 \multicolumn{2}{c||}{Consumption Loan} &  \multicolumn{2}{c|}{Cash Loan}\\
\cline{2-5} & Trade & Friendship  & Trade & Friendship \\
\hline
$\frac{br(\mathcal{N}_{y=1}^{(1)})}{br(\mathcal{N}_{y=0}^{(1)})}$ & 8.9 & 4.1 & 6.7 & 2.6 \\
\hline
$\frac{br(\mathcal{N}_{y=1}^{(2)})}{br(\mathcal{N}_{y=0}^{(1)})}$ & 3.36 & 1.75 & 2.43 & 1.52  \\
\hline
\end{tabular}
\label{tab:data_analysis1}
\end{table}

\begin{figure}
\centering
\subfigure[bad rate lift for consumption loan]{
\includegraphics[width=0.23\textwidth]{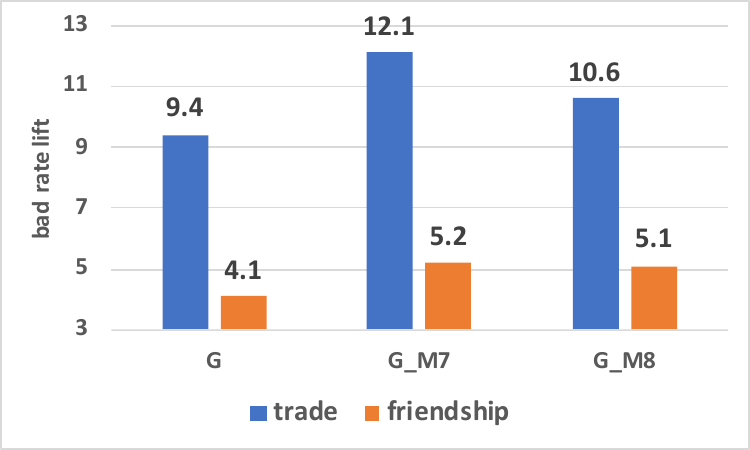}}
\subfigure[bad rate lift for cash loan]{
\includegraphics[width=0.23\textwidth]{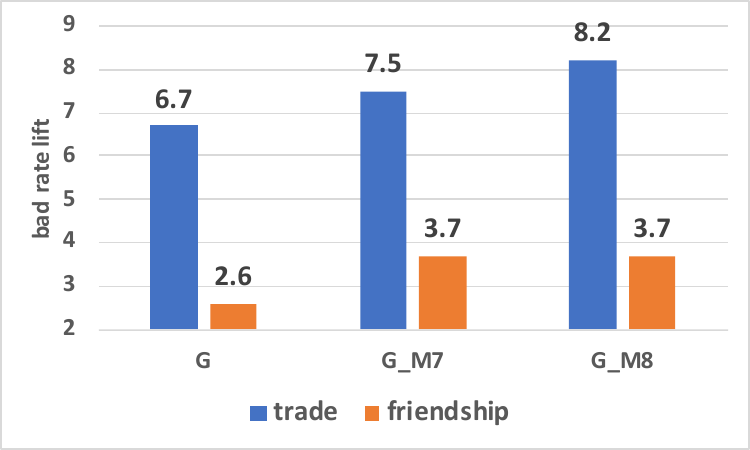}}
\subfigure[heterophily edge ratio for consumption loan]{
\includegraphics[width=0.23\textwidth]{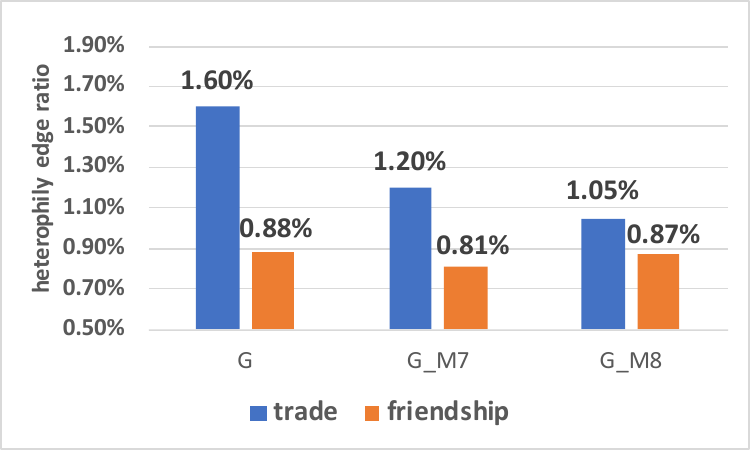}}
\subfigure[heterophily edge ratio for cash loan]{
\includegraphics[width=0.23\textwidth]{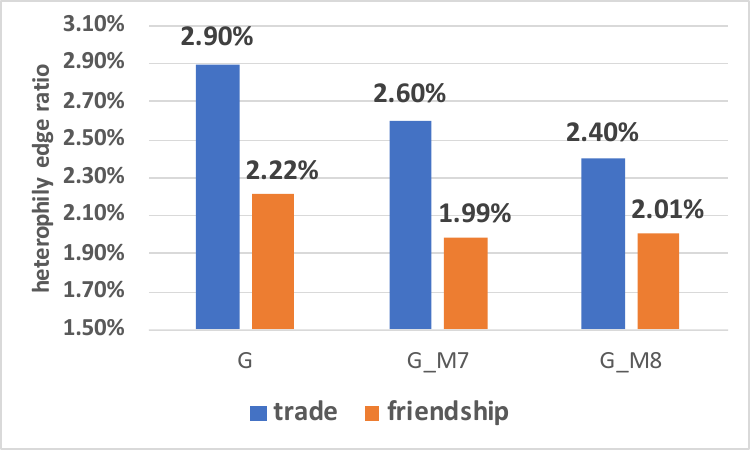}}
\caption{ The bad rate lift and heterophily edge ratio between the original social graph, denoted as G, and two types of motif-based graph, denoted as $G_{M_7}$ and $G_{M_8}$, on the two types of loan labels. }
\label{pic:data_analysis}
\end{figure}

However, both the first-order and second-order neighbors are lower-order graph structures at the level of paired users and edges. Going beyond the lower-order graph structure, we further look at the effect of the higher-order graph structure at the level of small subgraphs, i.e. motifs. We build the motif-based graphs based on the motif of $M_7$ and $M_8$, which can be regarded as a specific sub-graph of the original social graph. The details can be found in Section \ref{sec:motif}. We compare the bad rate lift and heterophily edge\footnote{An edge linking two nodes with different labels is defined as the heterophily edge.} ratio on the original graph and motif-based graph in Figure \ref{pic:data_analysis}. 

We find that, using different types of motif-based graphs can further increase the lift between the bad rate of default users' neighbors and normal users' neighbors, which demonstrates the higher-order structures can provide more discriminative information for financial default prediction. Furthermore, in Figure \ref{pic:data_analysis}(c-d), we find that the heterophily edge ratio in the motif-based graph will be greatly reduced. Many previous literatures have demonstrated that the heterophily edges usually have a bad effect on the learning process of GNNs \cite{bo2021beyond}. Therefore, incorporating the motif-based graph will enhance the homophily property of the original graph and thereby have the potential to obtain a better result. 

In summary, the aforementioned results all demonstrate that both the lower-order and higher-order graph structures contribute to financial default prediction. Then we will introduce how to model such graph structures to do accurate user default prediction. 

\section{Method}

\begin{figure*}[htb]
\centering
\includegraphics[width=0.9\textwidth]{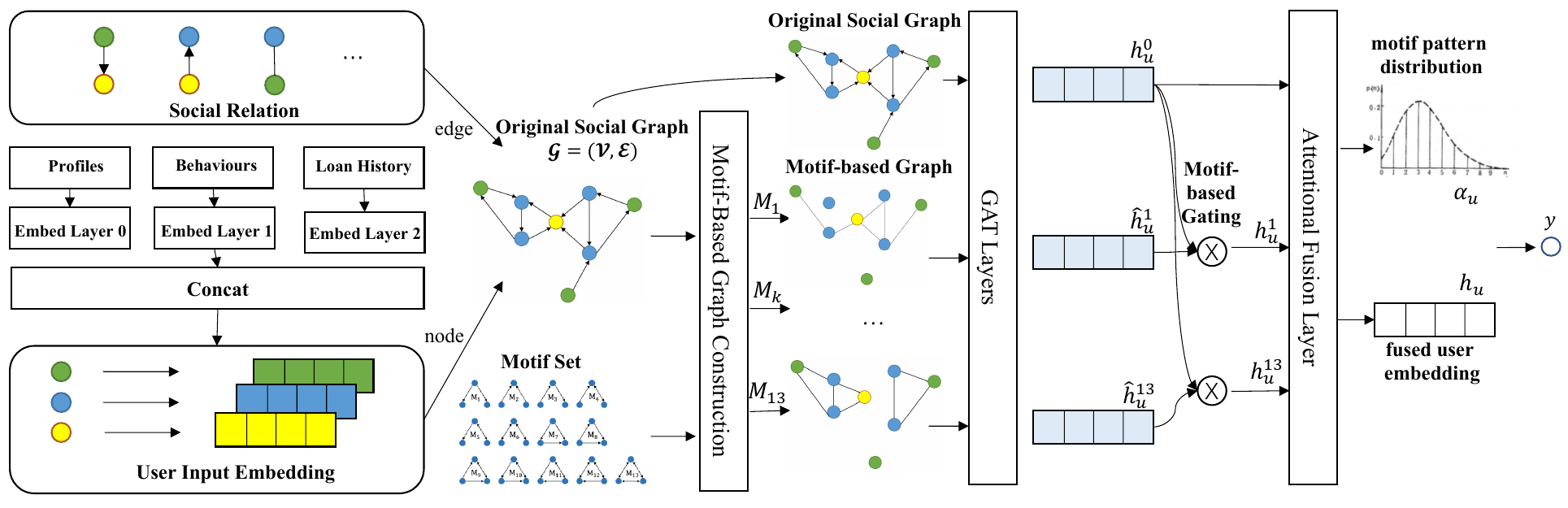}
\caption{ The framework of our proposed \methodshort. }
\label{pic:framework}
\end{figure*}

In this section, we will introduce our method, whose framework is shown in Figure \ref{pic:framework}. The whole model can be divided into four parts: (a) The input module consists of three separate embedding layers to encode the user's profiles, behaviors and loan history and form the user's input embeddings. (b) Motif-based Graph Construction Module generates multi-view motif-based graphs from different types of motifs, which explicitly retain various high-order graph structures for the following learning process. (c) Motif-wise Message Aggregation Module aggregates the neighbors from the multi-view graphs to obtain motif-wise node embeddings. (d) Motif-level Attentional Fusion Layer ensembles the multi-view motif-wise node embeddings for the financial default prediction. 

\subsection{The Input Module}
The Input Module consists of two parts: social graph construction and user input embedding generation. The building of the social graph has been introduced in Section \ref{sec:pf}. Note that for simplicity, we use the directed unweighted graph in this paper. We build the user input embedding from three aspects. The first is from the user's own demographics. The second is the user's behavior in the platform, which mainly includes some statistics about the log-in behavior, scan behavior, trade behavior and transfer behavior. The third part includes the user's historical loan information, like the loan amount, the loan frequency, the default times and so on. All the raw features are first quantified into discrete buckets, and then represented by the one-hot bucket ID. Then we use three individual embedding layers to do embedding-look-up on three parts of features and finally concatenate them to form the final user input embedding.

\subsection{Motif-based Graph Construction}
\label{sec:motif}
We first introduce the definition of motif as follows: 
\begin{myDef}
($l$-node Motif) A motif $M=(\mathcal{V}_m,\mathcal{E}_m)$ is a connected small graphs, where $|\mathcal{V}_m|=l$.
\end{myDef}

\begin{figure*}
\centering
\subfigure[$3$-node motif]{
\includegraphics[width=0.4\textwidth]{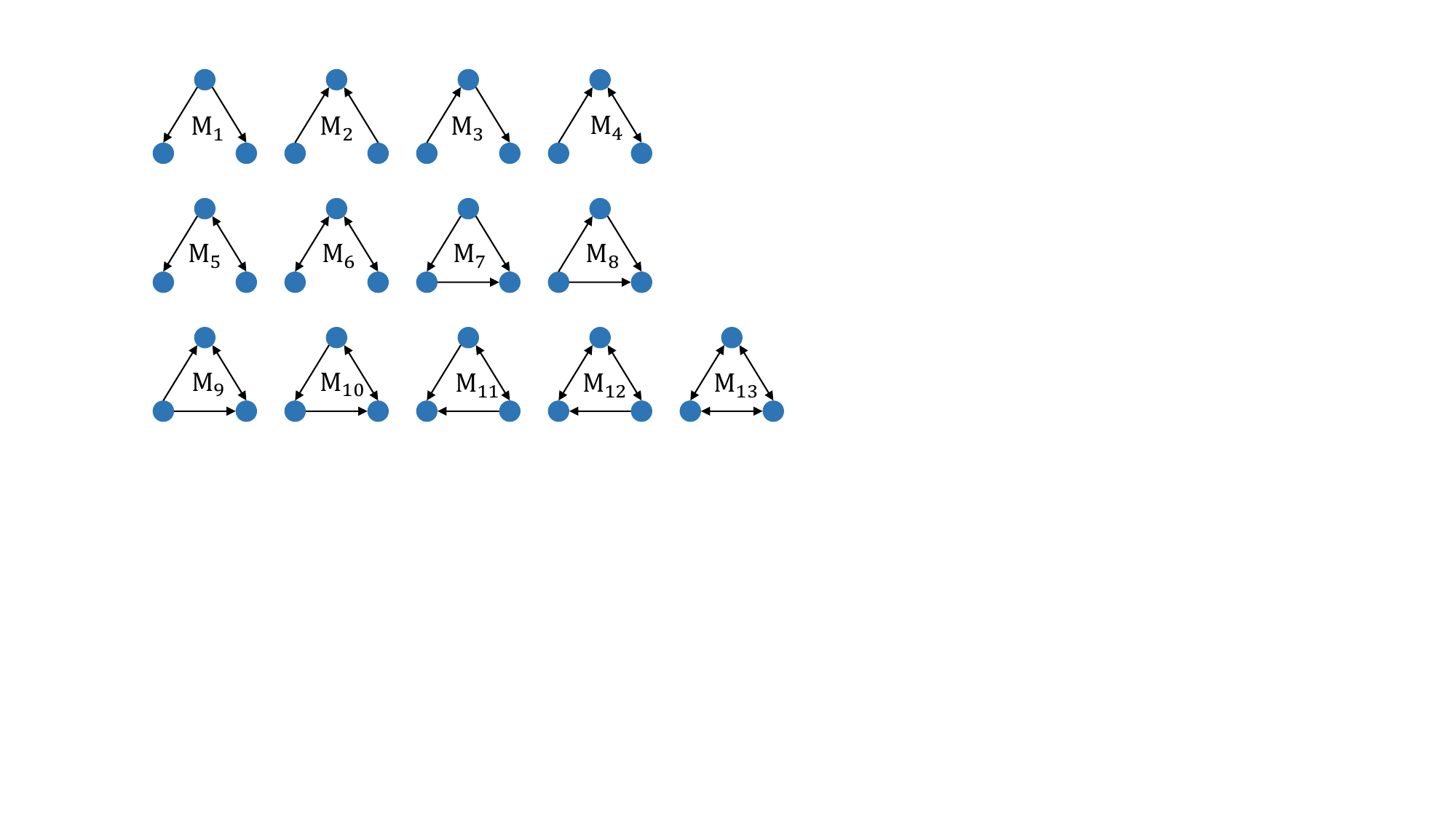}}
\subfigure[Construction of motif $M_8$-based graph]{
\includegraphics[width=0.5\textwidth]{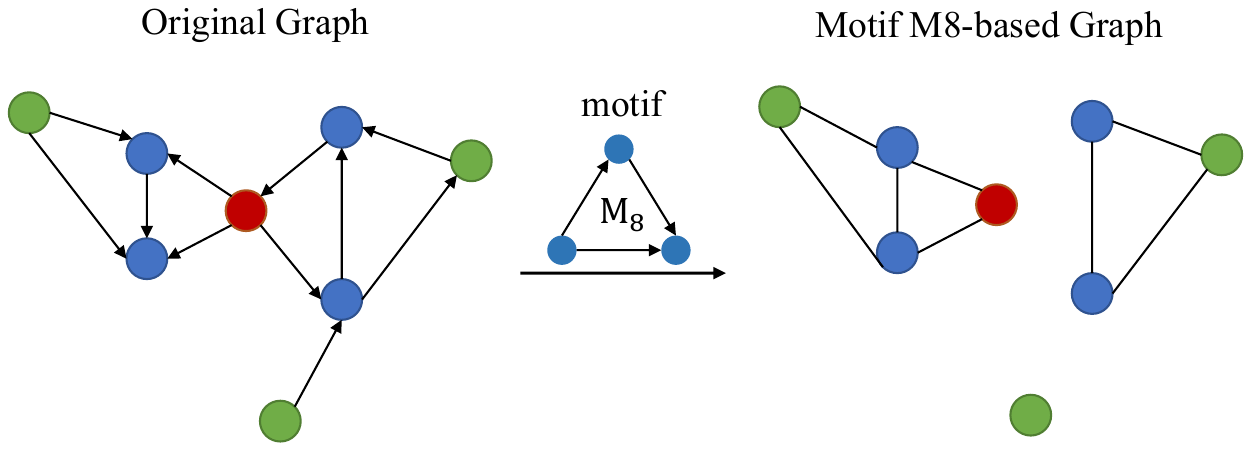}}
\caption{ The illustration of (a) all $3$-node motifs and (b) the construction process of motif $M_8$-based graph. }
\label{pic:base_motif}
\end{figure*}
We show all possible $3$-node motifs in Figure \ref{pic:base_motif}(a). $3$-node motifs are the most common and important building blocks for social graphs. Therefore in this work we only consider $3$-node motifs and the motif set is denoted as $\mathcal{M}=\{M_1,...,M_K\}$, where $K=13$.

Then we give the definition of motif instance as follows:
\begin{myDef}
(Motif Instance) Given the social graph $\mathcal{G}$ and an edge set $\mathcal{E}'$, we can obtain the subgraph $I=\mathcal{G}[\mathcal{E}']$. If $\mathcal{G}[\mathcal{E}']$ and a motif $M$ are isomorphic, then we say $I$ is a motif instance of motif $M$.
\end{myDef}

Note that a motif $M$ can have several motif instances given a graph. Each instance has a unique node set and two motif instances can share nodes. The set of all the motif instances of motif $M$ is denoted as $I(M)$ and all the instances containing node $u$ are denoted as $I_u(M)$.

At last, we define the motif-based adjacency matrix as follows: 
\begin{myDef}
(Motif-based Adjacency Matrix) Given a motif $M_k$ and the instance set $I(M_k)$, the motif-based adjacency matrix is defined as follows: 
\begin{equation}
\label{eq6}
(\mathbf{A}_k)_{ij}=\left\{
\begin{aligned}
1 &, i = j, \\
1 &, (u_i,u_j)\in I_u(M_k), \\
0 &, otherwise. 
\end{aligned}
\right.
\end{equation}
\end{myDef}

One natural question is how to generate the motif-based adjacency matrix. Theoretically, it has the time complexity of $O(n^K)$ for a motif with $K$ nodes, which is too consuming for the real applications. Actually, most real-world graphs are sparse. Following the works of \cite{benson2016higher,latapy2008main}, the motif instances of $3$-node motif can be found in $O(m^{1.5})$ time. 

Figure \ref{pic:base_motif}(b) shows an example of the motif graph construction. Generally speaking, the edges in the original graph that are isomorphic with the given motif are preserved in the motif-based graph, and the other edges are discarded. Therefore, several adjacent nodes will be disconnected in the motif-based graph and the connectivity of the motif-based graph is weakened, which is neglected by the previous works and also pose great challenges for graph learning. We will propose our solution in our model. 

\subsection{Motif-Wise Message Aggregation}

We denote the adjacency matrix of the original social graph as $\mathbf{A}_0$ and we combine all the adjacency matrix, including the original graph's and motif-based graphs', as $\mathcal{A}=\{\mathbf{A}_k\}_{k=0}^K$ to form the multi-view graph. Then for each adjacency matrix, we first utilize a graph attention network (GAT) to learn the node embeddings on each type of motif-based graph\footnote{For the simplicity of description, we also regard the original graph as a special kind of motif-based graph}. Specifically, we take the motif $M_k$ as an example. The motif $M_k$-based representation of node $u$ in the $l$-th layer is given by:
\begin{equation} \label{eq_hdl}
	\begin{gathered}
		 \mathbf{h}^{(l+1)}_{u,k} = tanh \left(\sum_{v \in \mathcal{N}_{k}(u) \cup \{u\}} 
		\boldsymbol{\alpha}(\mathbf{h}_{u,k}^{(l)},\mathbf{h}_{v,k}^{(l)})
		 \mathbf{h}^{(l)}_{v,k}  \mathbf{W}_{k}^{(l)} \right),
	\end{gathered}
\end{equation}
where $\mathcal{N}_{k}(u)$ denotes the set of user $u$'s neighbors in the motif $M_k$-based graph, and $\boldsymbol{\alpha}(\mathbf{h}_{u,k}^{(l)},\mathbf{h}_{v,k}^{(l)})$ is the attention function to measure the importance of user $i$ and user $j$, defined as:

\begin{equation} \label{eq_att}
	\begin{gathered}
		\boldsymbol{\alpha}(\mathbf{h}_{u,k}^{(l)},\mathbf{h}_{v,k}^{(l)}) = softmax(\mathbf{v}_k^{(l)}tanh(\mathbf{W}_{s,k}^{(l)}\mathbf{h}_{u,k}^{(l)} + \mathbf{W}_{d,k}^{(l)}\mathbf{h}_{v,k}^{(l)}))
	\end{gathered},
\end{equation}
where  ${\mathbf{W}_{s,k}}^{(l)}$ represents the weight for the source node, ${\mathbf{W}_{d,k}}^{(l)}$ represents the weight for the target node and $\mathbf{v}_k^{(l)}$ denotes a vector to map the representations to a value.

Then $L$ layers are stacked to obtain node representations $\mathbf{z}_{u,k}= \mathbf{h}^{(L)}_{u,k}$ for motif $M_k$. Previous works will directly aggregate the embedding of  $\mathbf{z}_{u,k}$ for each motif to form the final user representations \cite{wang2022graph,hu2022mbrep,liu2021motif,piao2021predicting}. However, these methods meet the following problem. The linkage between two nodes in the motif-based graph characterizes their higher-order relationship like the triangle relationships. It indicates a stronger bondage relationship between the pair of nodes, which is different from the direct link relation of the original graph and is essential to be captured by the model. On the other hand, many adjacent nodes in the original graph will be disconnected in the motif-based graph. For example, in the Cora dataset, in the $M_8$-based motif graph, less than $1$\% edges of the original graph are preserved. In this way, the connectivity of the motif-based graph is greatly weakened. Considering this, the representation ability of the motif-based representation $\mathbf{z}_{u,k}$ is not that good.

Considering that the connectivity of the original graph is stronger, we hope to resort to the node embedding of the original graph $\mathbf{z}_{u,0}$ for help but still focus on capturing the higher-order graph structure. To achieve this, we propose a gated unit to distill the node embedding of the original graph. The basic intuition is that we build a gate according to the higher-order structural information to learn which part of the information learned by the lower-order structure should be strengthened. Specifically, we define a non-linear gated function $f_{GATE}^k(\cdot)$ for motif $M_k$ to modulate the $\mathbf{z}_{u,0}$ at a feature-wise granularity through re-weighting, defined by:
\begin{equation} \label{eq_att2}
	\hat{\mathbf{h}}_u^k=f_{GATE}^k(\mathbf{z}_{u,0})=\mathbf{z}_{u,0} \odot \sigma(\mathbf{W}_g^k \mathbf{z}_{u,k}+\mathbf{b}_g^k),
\end{equation}
where $\mathbf{W}_g^k$ and $\mathbf{b}_g^k$ are learnable parameters, $\odot$ denotes the element-wise product operation and $\sigma(\cdot)$ is the sigmoid non-linear function. The gating mechanism effectively serves as a skip-connection which focuses on preserving the higher-order information from the node embedding learned from the lower-order graph structure.

At last, we concatenate $\hat{\mathbf{h}}_u^k$ and $\mathbf{z}_{u,k}$ to form the final motif-wise user representations, given by $\mathbf{h}_u^k=\hat{\mathbf{h}}_u^k || \mathbf{z}_{u,k}$.

\subsection{Motif-level Attentional Fusion Layer}
With the multi-view motif-wise user representations, we aim to aggregate them to obtain the fused user representations. Considering that different users are likely to have diverse preferences over different higher-order structures from multiple views. We thus devise a motif-level attention mechanism to capture user preferences towards motifs. In detail, the attention score for each motif is defined as:
\begin{equation}
    s(\mathbf{h}_u^k) = \tanh{(\mathbf{w}_a^k \mathbf{h}_u^k)},
\small
\end{equation}
where $\mathbf{w}_a^t$ denotes the learnable vector to map the motif-vise embedding into a score. 

With the attention score, we use the softmax to get the attention coefficients for each motif:
\begin{equation}
    \alpha_u^k = \frac{\exp{(s(\mathbf{h}_u^k))}}{\sum_{t\in \{0,1,...,K\}}{\exp{(s(\mathbf{h}_u^t))}}}. 
\end{equation}

Then, the fused user embedding is defined as the sum over all the motif-wise user representations: 
\begin{equation}
    \mathbf{h}_u = \sum_{i=0}^{K}{ (\alpha_k\cdot \mathbf{h}_u^k)}. 
\label{eqn:user_embedding}
\end{equation}

\subsection{Curriculum Learning-based Loss Functions}
As we have stated before, the motif distributions are unbalanced. Some motifs are relatively common but some are uncommon. As a result, the structure of each user, a combination of lower-order structure and higher-order structure based on different motifs, is highly unbalanced. Some users may carry uncommon motif-based structures, which are difficult to be captured and be underestimated. While most users carry common motif patterns, which makes the model skew to these samples. Therefore, the structure unbalance may result in a great bias issue, which is still an unsolved problem for existing works. 

To overcome the problem, we propose a motif-based curriculum learning mechanism to do sample re-weighting. Specifically, we use the vector $\boldsymbol{\alpha}_u=\{\alpha_u^k\}_{k=1,...,K}$ to denote the attention weights for user $u$ over the motif set. It characterizes the distribution of user preferences on different types of motifs. If the distribution diverges from other users, it may indicate that the user contains uncommon motif patterns. In this way, we should emphasize the learning of the sample. To achieve this, we quantify the sample weight $\boldsymbol{\beta}_u$ by the deviation of $\boldsymbol{\alpha}_u$ from the mean distribution of the labeled nodes, given by
\begin{equation}
\left\{
\begin{aligned}
\beta_u &= \frac{exp(||\boldsymbol{\alpha}_u-\boldsymbol{\mu}||^2)}{\sum_{v\in \mathcal{V}_L}{exp(||\boldsymbol{\alpha}_v-\boldsymbol{\mu}||^2)}} \\
\boldsymbol{\mu} &= \frac{1}{|\mathcal{V}_L|} \sum_{v\in \mathcal{V}_L}{\boldsymbol{\alpha}_v}
\end{aligned}
\right.,
\end{equation}

The sample weight is normalized to $0<\beta_u<1$ over the values of all labeled users using a softmax function. Then we use the multi-perceptron layers (MLP) to map the fused user embedding $\mathbf{h}_u$ into the predicted default probability $\hat{y}_u$ and define our final supervised loss as:
\begin{equation}
\mathcal{L} = -\frac{1}{|\mathcal{V}_L|}\sum\limits_{u\in \mathcal{V}_L}{\beta_u (\hat{y}_u\log{(y_u)}+(1-\hat{y}_u)\log{(1-y_u)})}+L_{reg},
\label{eqn:loss}
\end{equation}
where $y_u \in \{0,1\}$ is the ground-truth label for the user $u$ and $L_{reg}$ is the $L_2$-regularizer of the model parameters . 

In contrast to the common supervised cross-entropy loss for binary classification, our loss function induces a training curriculum that progressively focuses on samples with uncommon motif patterns, which has the potential to learn more comprehensive motif patterns to get a better result. 

\subsection{Discussions}
\subsubsection{Complexity Analysis}
As we have stated, the time complexity to obtain a motif-based graph is $O(m^{1.5})$. As for the graph learning, it is similar to the common GCNs learning process, whose time complexity is $O(m)$ for each graph. The following fusing and curriculum learning process has the time complexity of $O(n)$, since each node now is independent. In sum, the overall time complexity of our model is $O(m^{1.5})$, where each motif-based graph can be processed parallelly. 

\subsubsection{Generalization}
Another interesting question is whether the method can be applied to $4$ or even more node motifs. Actually, the method can work for any type of motif by generating the motif-based graphs according to the given motif. Then the following learning process is the same. But in practice, we do not use the motif with more than $3$ nodes. On the one hand, the time complexity for generating a $4$-node motif is high. On the other hand, the motif-based graphs for a $4$-node motif are very sparse, which poses great challenges for graph-based methods to learn.

\section{Experiment}
\subsection{Experiment Setup}
\subsubsection{Dataset} 
To demonstrate the effectiveness of our proposed method, \methodshort, we conduct extensive experiments on one public dataset: Cora, and two industrial datasets denoted as ConsumeLn and CashLn. ConsumeLn and CashLn are collected from one of China's fin-tech platforms, where nodes represent the users and edges represent their relationships, including the friendship, transfer and trade relations. Note that we do not differentiate different types of relations since it is not the focus of our paper and we leave it as the future work. Each user can be characterized by three parts of features: $82$-dimension profiles, $536$-dimension behavior history and $241$-dimension loan history. Each user has two default labels each month: one label for a consumption-loan product ConsumeLn and one label for a cash-loan product CashLn. A one-year-long dataset in $2021$ is used for the experiment. The first half year is used for training, the third quarter is used for validation and the last quarter is used for testing. Note that all the above data are definitely authorized by the users and be anonymous. 

Since there is no public dataset for financial default prediction, to demonstrate the generality of our proposed method we also give the performance on Cora. Cora is a very common benchmark dataset, where nodes correspond to papers and edges correspond to their citation relationship. Each paper is associated with a bag-of-words feature vector and the task is to classify the paper into one of $7$ research topics. 

\begin{table}
\caption{Statistics of our datasets. }
			\begin{tabular}{cccccc}
				\textbf{Datasets}&$|\mathcal{V}|$&$|\mathcal{E}|$&$|V_L|$&$d_{nd}$\\
				\hline
				\textbf{ConsumeLn}  &  1B  &  3.2B  & 50M   & 879  \\
				\textbf{CashLn}  &  80M  &  230M  & 3.7M   & 852  \\
                \textbf{Cora}  &  2485  &  5069  & 2485   & 1433  \\
		\end{tabular}
\label{tab:dataset}
\end{table}
The details of three datasets can be found in Table \ref{tab:dataset}.

\subsubsection{Comparing Methods}
In order to validate the performance of our model, we compare it with four lines of state-of-the-art methods.
\begin{itemize}
    \item Tree-based methods: We use Random Forest \cite{breiman2001random} and XGBoost as representatives of tree-based methods. 
    \item Proximity-based graph embedding method: This type of method only encodes the graph topology and aims to make topologically nearby nodes have similar embeddings. We use Node2Vec \cite{grover2016node2vec} and motif2vec \cite{dareddy2019motif2vec} as representative methods. 
    \item Graph Neural Networks: We choose GCN \cite{kipf2016semi}, GAT \cite{velivckovic2017graph}, GraphSAGE \cite{hamilton2017inductive} as representatives of genenral GNNs.
    \item GNN-based Financial Default Prediction: GeniePath \cite{liu2019geniepath}, MvMoe \cite{liang2021credit} and IGNN \cite{wang2019semi} are GNN-based methods specifically designed for financial default prediction. 
    \item Motif-based Methods: MCN firstly generates the motif-based graphs and uses GNNs to generate embedding on each graph. At last, it proposes a motif-level attention mechanism to aggregate the node embedding generated from each motif-based graph. 
\end{itemize}

\subsubsection{Metrics and Reproducibility}
For Cora, we adopt a widely-accepted metric: Accuracy as the evaluation metric. For ConsumeLn and CashLn, since the dataset is highly unbalanced regarding the labels, we adopt AUC and KS (a widely used metric for default prediction \cite{massey1951kolmogorov}) as the evaluation metrics. 

For Cora, we implement the experiment with Pytorch, where we use two layers of GAT with a dimension size of $256$. The model is trained for a maximum of $60$ epochs with a batch size of $256$ with the Adam optimizer. The learning rate is tuned in the range $\{10^{-3},5*10^{-3},10^{-2}\}$. For ConsumeLn and CashLn, we implement the experiments with Tensorflow because it has been deployed online and has to be fit with our industrial production settings. The output size of embedding layers in the input module are $128$ for three parts of user features. We also use two layers of GNN with a dimension size of $256$. The model is trained for a maximum of $3$ epochs with a batch size of 512 with the Adam optimizer. The learning rate is tuned in the range $\{10^{-5},5*10^{-4},10^{-4}\}$. Following prior literature, we include all thirteen $3$-node motifs as candidates. Note that due to the size of industrial dataset, it is unrealistic to load all the samples into the memory. For the calculation of $\mu$ in Eq. 8, we only consider the labeled data in the current batch. Note that we will run each algorithm $5$ times to report the averaged results. 

\subsection{Overall Performance}
\subsubsection{Results on Cora}

\begin{table}[htb]
\centering
\caption{ Test Accuracy on Cora with the training ratio of 20\%, 40\% and 60\%. X denotes the node features and $\mathcal{G}$ denotes the graph topology.  }
\begin{tabular}{|c|c c||c c c|}
\hline
 &
 \multicolumn{2}{c||}{Data} &  \multicolumn{3}{c|}{Cora}\\
\hline
Training Ratio & X & $\mathcal{G}$  & 20\% & 40\% & 60\%  \\
\hline
\multicolumn{6}{|c|}{Tree-based Method} \\
RF & \checkmark & & 0.551 & 0.731 & 0.741  \\
XGBoost & \checkmark & & 0.618 & 0.742 & 0.749  \\
\hline
\multicolumn{6}{|c|}{Proximity-Based Graph Embedding} \\
Node2Vec &  & \checkmark & 0.757 & 0.761 & 0.776 \\
Motif2Vec &  & \checkmark & 0.790 & 0.792 & 0.798 \\
\hline
\multicolumn{6}{|c|}{Graph Neural Network} \\
GCN & \checkmark & \checkmark & 0.816 & 0.820 & 0.830\\
GAT & \checkmark & \checkmark & 0.809 & 0.814 & 0.818 \\
GraphSAGE & \checkmark & \checkmark & 0.813 & 0.835 & 0.842 \\
\hline
\multicolumn{6}{|c|}{GNN-based Financial Default Prediction} \\
GeniePath & \checkmark & \checkmark & 0.810 & 0.832 & 0.841 \\
\hline
\multicolumn{6}{|c|}{Motif-based GNN model} \\
MCN & \checkmark & \checkmark & 0.811 &  0.824 & 0.831  \\
\methodshort & \checkmark & \checkmark& \textbf{0.832} & \textbf{0.859} & \textbf{0.865} \\
\hline
\end{tabular}
\label{tab:cora_performance}
\end{table}

To get a comprehensive result, we evaluate different methods on the training ratio of 20\%, 40\% and 60\% and test ratio of 20\%. The results are shown in Table \ref{tab:cora_performance}. We find that: 
\begin{itemize}
    \item RF and XGBoost performs worse than Node2vec and Motif2vec, which demonstrates that the graph topology, especially motif-level information, is very important in this dataset. 
    \item GNN-based methods outperform tree-based method and graph embedding method, which demonstrates that integrating the node attributes and graph topology are essential.
    \item MCN performs worse than GraphSAGE, which demonstrates that simply integrating the motif information will not improve the performance necessarily.
    \item Our propose method \methodshort{} outperforms all the comparing methods, which demonstrates that our method can better capture the lower-level (edge-level) and higher-level (motif-level) structure information for a better graph-based learning. 
\end{itemize}

\subsubsection{Results on Industrial Datasets}

\begin{table}[htb]
\centering
\caption{ Test Accuracy on ConsumeLn and CashLn. X denotes the node features and $\mathcal{G}$ denotes the graph topology.  }
\begin{tabular}{|c|c c||c c|c c|}
\hline
 &
 \multicolumn{2}{c||}{Data} &  \multicolumn{2}{c|}{ConsumeLn}  &  \multicolumn{2}{c|}{CashLn}\\
\hline
Evaluation Metric & X & $\mathcal{G}$  & AUC & KS & AUC & KS  \\
\hline
\multicolumn{7}{|c|}{Tree-based Method} \\
RF & \checkmark & & 0.782 & 0.435 & 0.696 & 0.282   \\
XGBoost & \checkmark & & 0.781 & 0.434 & 0.696 & 0.283   \\
RF-Graph & \checkmark & & 0.791 & 0.440 & 0.700 & 0.287   \\
XGBoost-Graph & \checkmark & & 0.793 & 0.442 & 0.700 & 0.288   \\
RF-Motif & \checkmark & & 0.796 & 0.443 & 0.703 & 0.292   \\
XGBoost-Motif & \checkmark & & 0.797 & 0.445 & 0.702 & 0.291   \\
\hline
\multicolumn{7}{|c|}{Proximity-Based Graph Embedding} \\
Node2Vec &  & \checkmark & 0.640 & 0.203 & 0.605 & 0.141 \\
Motif2Vec &  & \checkmark & 0.652 & 0.209 & 0.611 & 0.152 \\
Node2Vec+features & \checkmark & \checkmark & 0.784 & 0.437 & 0.697 & 0.282 \\
Motif2Vec+features & \checkmark & \checkmark & 0.789 & 0.439 & 0.699 & 0.287  \\
\hline
\multicolumn{7}{|c|}{Graph Neural Network} \\
GCN & \checkmark & \checkmark & 0.808 & 0.464 & 0.731 & 0.335 \\
GAT & \checkmark & \checkmark & 0.814 & 0.473 & 0.736 & 0.343 \\
GraphSAGE & \checkmark & \checkmark & 0.810 & 0.468 & 0.732 & 0.338 \\
\hline
\multicolumn{7}{|c|}{GNN-based Financial Default Prediction} \\
GeniePath & \checkmark & \checkmark & 0.814 & 0.475 & 0.737 & 0.345 \\
MvMoe & \checkmark & \checkmark & 0.810 & 0.469 & 0.727 & 0.329 \\
IGNN & \checkmark & \checkmark & 0.816 & 0.478 & 0.738 & 0.348 \\
\hline
\multicolumn{7}{|c|}{Motif-based GNN model} \\
MCN & \checkmark & \checkmark & 0.818 & 0.480 & 0.738 & 0.347  \\
\methodshort & \checkmark & \checkmark& \textbf{0.823} & \textbf{0.495} & \textbf{0.745} & \textbf{0.360} \\
\hline
\end{tabular}
\label{tab:loan_performance}
\end{table}

We report the results on our industrial datasets ConsumeLn and CashLn in Table \ref{tab:loan_performance}. Note that RF(XGBoost)-Graph is based on input user features and mined graph-based features. RF(XGBoost)-Motif is based on input user features, graph-based features and motif-based features. From the table, we have the following observations and conclusions:
\begin{itemize}
    \item Our model constantly outperforms all baseline methods on all the metrics, which demonstrates the effectiveness of our method. To note that although the absolute performance gain is not that much compared with the best-performing baseline method, it can bring a great profit for real business. 
    \item Comparing the performance of RF(XGBoost), RF(XGBoost)-Graph and RF(XGBoost)-Motif, we find out utilizing graph structure, especially higher-order graph structure captured by motifs is beneficial. 
    \item We find that Node2Vec and Motfi2Vec perform badly in this dataset. The reason is that for default prediction, the user's attributes are much more important than the graph topology. Combining user attributes and graph structure in a sophisticated way is very important for financial default prediction. 
    \item We find that motif-based methods MCN and MotifGNN outperform GNN-based financial default methods, which demonstrates that capturing the motif-based relations is essential and beneficial for the task of financial default prediction. 
    \item Comparing MCN and MotifGNN, we find that our proposed method can use the motifs to capture the higher-order graph structure more effectively.
\end{itemize}

\begin{figure*}
\centering
\subfigure[Cora]{
\includegraphics[width=0.3\textwidth]{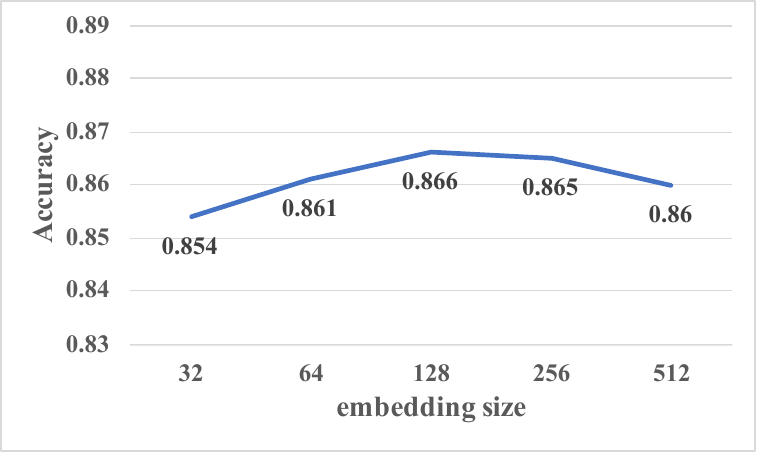}}
\subfigure[ConsumeLn]{
\includegraphics[width=0.3\textwidth]{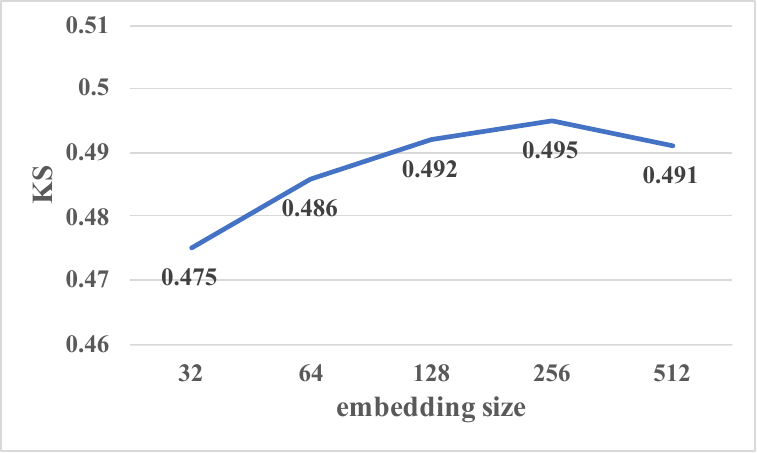}}
\subfigure[CashLn]{
\includegraphics[width=0.3\textwidth]{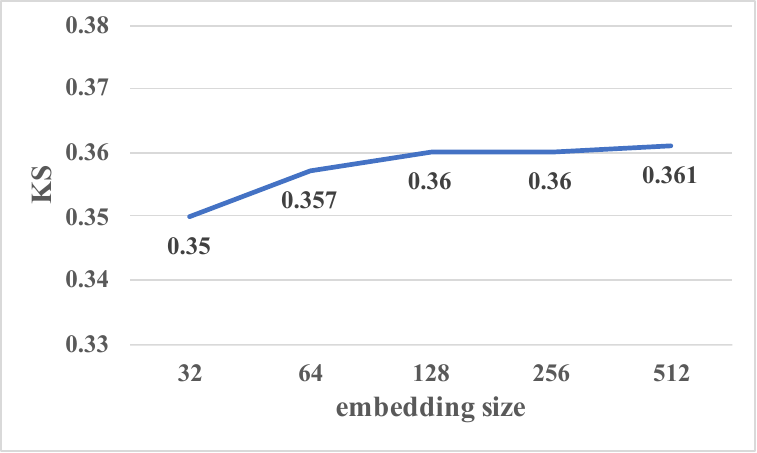}}
\caption{ Parameter Sensitivity Analysis on (a) Cora (b) ConsumeLn (c) CashLn. }
\label{pic:sensitivity}
\end{figure*}

\subsection{Ablation Study}
\begin{figure}
\centering
\subfigure[KS on ConsumeLn]{
\includegraphics[width=0.23\textwidth]{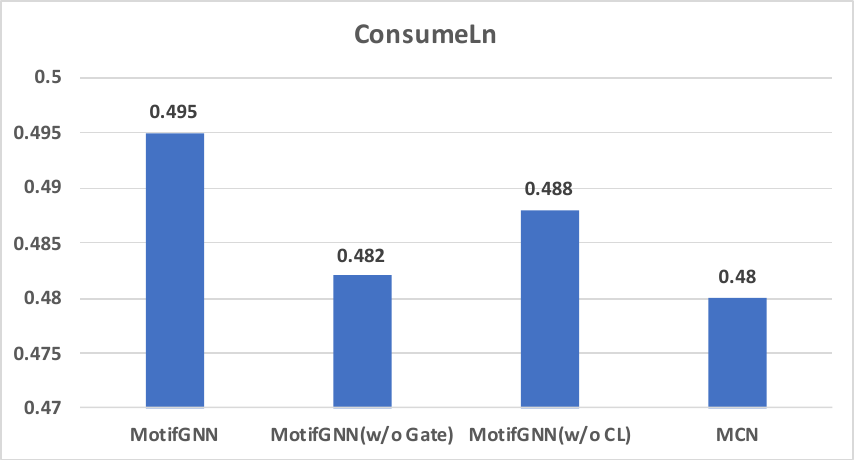}}
\subfigure[KS on CashLn]{
\includegraphics[width=0.23\textwidth]{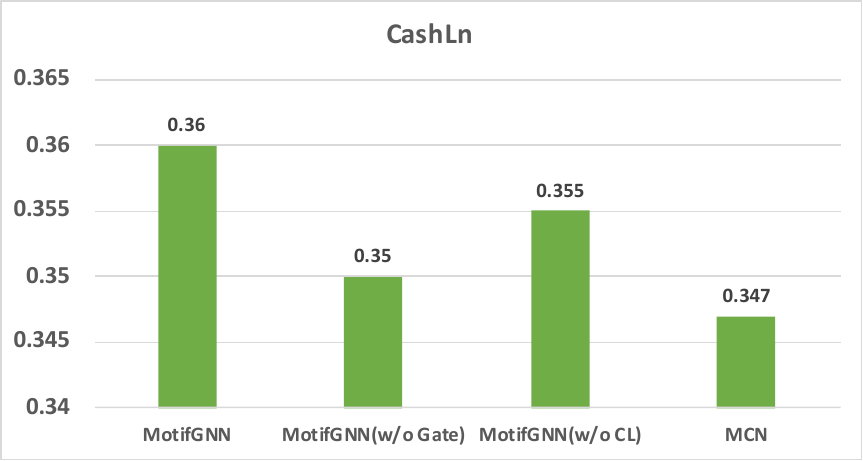}}
\caption{ Ablation Study on ConsumeLn and CashLn. }
\label{pic:ablation_study}
\end{figure}

Our proposed method includes two key components: the motif-based gate and the curriculum learning. To analyze the effect of each component, we conduct the ablation study on the ConsumeLn and CashLn datasets. The variants of our method include \methodshort without the motif-based gate (\methodshort(w/o MoG)) and 
\methodshort without the curriculum learning (\methodshort(w/o CL)). The results are shown in Figure \ref{pic:ablation_study}. We find that both two components contribute to performance improvement. Interestingly, we find that the improvement of \methodshort(w/o Gate) over MCN is not obvious but the improvement of \methodshort over \methodshort(w/o CL) is much larger. The reason is that the quality of curriculum learning is greatly affected by the quality of motif-level attention. Without the proposed motif-based gating mechanism, the model cannot learn very effective knowledge from the motif-based graphs due to their weak connectivity. In this way, the motif-based attentions are underestimated for some motifs, which makes the curriculum learning not effective enough. This result demonstrates that these two components are coupled with each other. 

\subsection{Analysis about Motifs}
\begin{figure}[htb]
\centering
\includegraphics[width=0.45\textwidth]{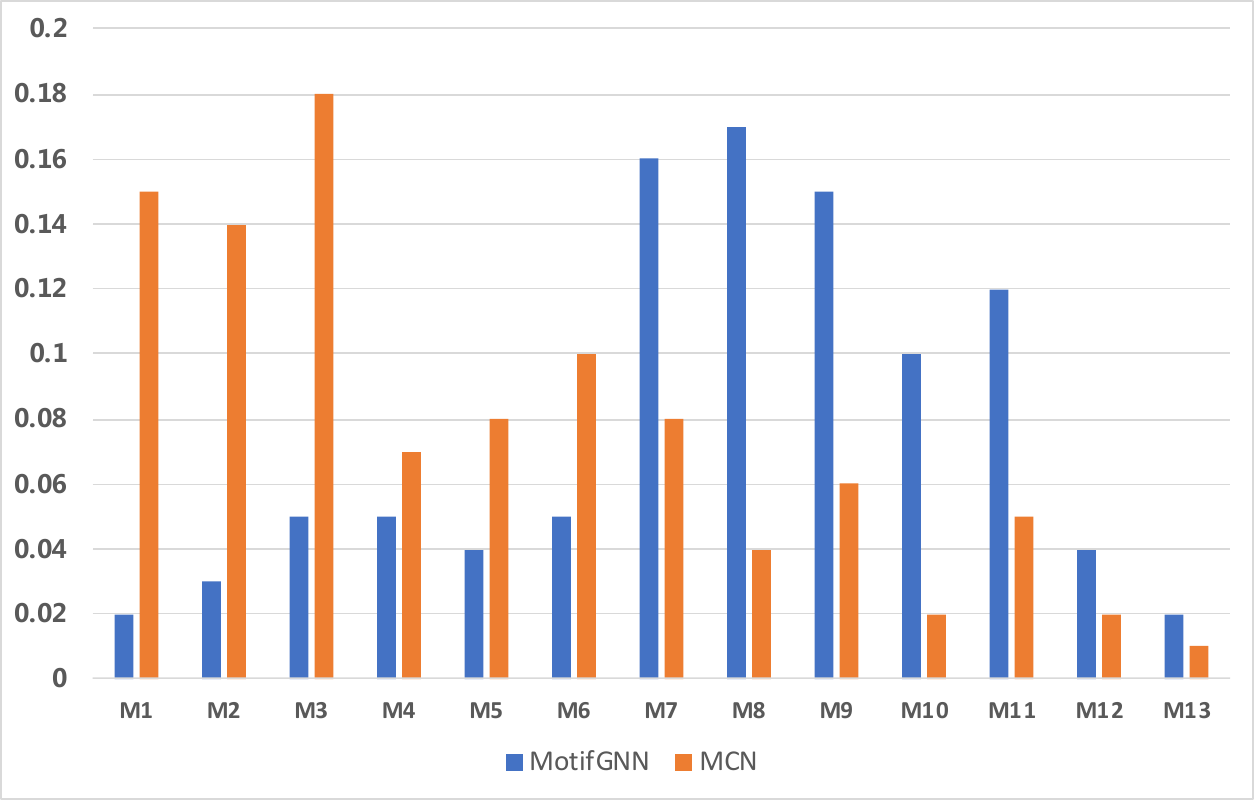}
\caption{ Distribution of attention coefficients among different types of motif-based graphs on our proposed method \methodshort and the state-of-the-art method MCN in ConsumeLn.  }
\label{pic:attention}
\end{figure}

We make an in-depth analysis regarding the effect of motifs in this section. We give the distribution of attention coefficients among different types of motif-based graphs in Figure \ref{pic:attention}. It is worth noting that MCN is also the motif-based GNN method, which directly uses the embedding $\hat{\mathbf{h}}_u^k$ to obtain the fused user embedding. From Figure \ref{pic:attention}, we have the following observations and analysis: 
\begin{itemize}
    \item We find that varying attention weights are assigned to different motifs, which demonstrates that the effects of different motifs are different. 
    \item MCN more focuses on the motifs with fewer edges such as $M_0$,$M_1$ and $M_2$. But for other motifs, the weights are very small. The reason is that the motif-based graphs of $M_7$ to $M_{13}$ have much fewer edges than the motif-based graphs of $M_1$ to $M_3$, thereby MCN cannot learn effective information within these motif-based graphs with low connectivity. 
    \item Our method more focuses on the motifs with triangle relationships, which is in correspondence with many works in social analysis that the triangle relationship is a very strong bondage between users. Thus the users with triangle relationships have strong homophily regarding their credit risk, which is discriminative information for the model. Furthermore, it also demonstrates that although some motif-based graphs are with relatively weak connectivity, our proposed model can still extract the useful information from them.
\end{itemize}

\subsection{Parameter Sensitivity}
We report the parameter sensitivity analysis of \methodshort{} in terms of the embedding size in Figure \ref{pic:sensitivity}. We find that when the embedding size is too small like $32$, the performance drops quickly. The reason is that in this case, the learning ability of the model is limited to learn informative information from data. Then with the increase of the embedding size, the accuracy will increase and is stable in general. Then if the embedding size continuously increases, it may cause overfitting to the model and the accuracy may drop. 

\section{Conclusion}
In this paper, we investigate the study of default prediction on online financial platforms. We propose a Motif-preserving Graph Neural Network with Curriculum Learning to learn the lower-order and higher-order graph structure. In particular, we propose a motif-based gating mechanism to effectively leverage the higher-order structure from the motif-based graph. Then we define a metric to measure the deviation of the user's motif distributions from the averaged distribution and accordingly design a curriculum learning method to make the model progressively focus on the samples with uncommon motif distributions. Experimental results on a public dataset and two industrial datasets demonstrate the superiority of our method. Future work will focus on investigating the effect of more complex motif patterns like temporal motifs and heterogeneous motifs on financial default prediction.

\bibliographystyle{unsrt}
\balance
\bibliography{sample-base}

\end{document}